\pdfoutput=1
\documentclass[conference,10pt]{IEEEtran}
\usepackage[utf8]{inputenc}
\usepackage[english]{babel}
\usepackage{array}
\usepackage{multirow}
\usepackage{graphicx}
\usepackage[hyphens]{url}
\usepackage[
comma,        
numbers,       
sort&compress,         
]{natbib} 
\usepackage{tikz}
\usetikzlibrary{fit, arrows, calc, positioning, backgrounds, shapes.misc, shapes.geometric}
\usepackage{subcaption}
\usepackage{color}
\usepackage{listings}
\usepackage{environ}
\usepackage{siunitx}
\usepackage{hyperref}
\usepackage[]{algorithm2e}
\usepackage{bytefield}
\usepackage{listings}
\usepackage{mathtools}
\usepackage{placeins}
\usepackage{tabularx, ragged2e}
\usepackage[most]{tcolorbox}
\let\labelindent\relax \usepackage{enumitem}
\usepackage{booktabs}
\usepackage{float}
\usepackage{newfloat}
\usepackage{booktabs}
\usepackage{glossaries}
\usepackage{cleveref}
\usepackage[absolute]{textpos}
\usepackage{flushend}
\usepackage{newfloat}

\newcolumntype{C}{>{\Centering\arraybackslash}X}

\newcommand{\Requester}{Requester}
\newcommand{\Provider}{Provider}
\newcommand{\Backend}{Backend}
\newcommand{\Requesters}{Requesters}
\newcommand{\Providers}{Providers}

\newtcolorbox{protocolbox}{%
     enhanced, 
     unbreakable, 
     arc=0pt, 
     outer arc=0pt, 
     leftrule=0mm, 
     rightrule=0mm,
     colback=white,
     right=1mm,
     left=1mm
}

\newcounter{protocol}
\newenvironment{protocol}[1]
{ \par\smallskip\par\noindent
  \begin{protocolbox}
  \refstepcounter{protocol}%
  \textbf{Protocol \theprotocol} #1 \par\vspace{-\parskip}%
  \begin{enumerate}[label=(\alph*)]
  \itemsep0em
}{
  \end{enumerate}   
  \end{protocolbox}
  \par\smallskip\par
}

\DeclareFloatingEnvironment[
   fileext=frm,
   name=Listing
]{listing}

\DeclareCaptionSubType{listing}

\TPMargin{5pt}

\newcommand{\Title}{Protecting RESTful IoT Devices from Battery Exhaustion DoS Attacks}
\title{\Title \\[-2.0ex]}

\author{\IEEEauthorblockN{Stefan Hristozov\IEEEauthorrefmark{1},
Manuel Huber\IEEEauthorrefmark{1},
Georg Sigl\IEEEauthorrefmark{2}}
\IEEEauthorblockA{\IEEEauthorrefmark{1}Fraunhofer AISEC -- 
 \{stefan.hristozov, manuel.huber\}@aisec.fraunhofer.de }
\IEEEauthorblockA{\IEEEauthorrefmark{2}Technical University of Munich --
 sigl@tum.de}}

\hypersetup{
  colorlinks=true, linkcolor=black, citecolor=black, urlcolor=black,
  pdftitle={\Title}
}

\RequirePackage[%
  babel,%
  final,%
  expansion=alltext,%
  protrusion=alltext-nott]{microtype}%

\newacronym{ble}{BLE}{Bluetooth Low Energy}
\newacronym{coap}{CoAP}{Constrained Application Protocol}
\newacronym{dos}{DoS}{Denial of Service} 
\newacronym{ddos}{DDoS}{Distributed Denial of Service} 
\newacronym{ewma}{EWMA}{Exponentially Weighted Moving Average} 
\newacronym{rtls}{RTLS}{Real Time Locating Systems} 
\newacronym{ca}{CA}{Certification Authority} 
\newacronym{pki}{PKI}{Public Key Infrastructure}

\glsdisablehyper %

\begin{document}

\maketitle

\begin{abstract}
Many IoT use cases involve constrained battery-powered devices offering services in a RESTful manner to their communication partners. Such services may involve, e.g., costly computations or actuator/sensor usage, which may have significant influence on the power consumption of the service \Providers{}.
Remote attackers may excessively use those services in order to exhaust the \Providers' batteries, which is a form of a \gls{dos} attack.
Previous work proposed solutions based on lightweight symmetric authentication.
These solutions scale poorly due to requiring pre-shared keys and do not provide protection against compromised service \Requesters{}.
In contrast, we consider more powerful attackers even capable of compromising legit \Requesters{}. 
We propose a method that combines attacker detection and throttling, conducted by a third trusted \Backend{}, with a lightweight authentication protocol. 
For attacker detection and throttling, we propose a novel approach using rate limitation algorithms.
In addition,
we propose and formally verify two authentication protocols suitable for different, widely used IoT network topologies. 
Our protocols ensure service availability for benign \Requesters{} even if \Providers{} are under a battery exhaustion attack.
The protocols do neither require pre-shared keys between \Requesters{} and \Providers{}, nor the usage of asymmetric cryptography and public key infrastructures on the \Provider{}.
This makes our protocols suitable for a variety of  IoT deployments involving constrained devices and constrained networks.
We demonstrate the feasibility of our method through a simulation and a proof of concept implementation. 
\end{abstract} 

\section{Introduction}
The charge of battery-powered IoT devices presents an attractive target for DoS attacks.
For such devices, battery exhaustion DoS attacks are especially damaging because their effect does not recede after the attack.
In case of battery depletion, the device can only be reinstated if the battery is changed or recharged. In IoT deployments with a large number of possibly spatially dispersed battery-powered devices, this may cause tremendous effort, financial costs, and long downtimes of business or safety critical applications.

The threat model in this paper considers a malicious \Requester{} excessively making use of a \Provider's power-intensive application-layer services which may involve, e.g., costly computations, actuator/sensor usage or transmission of a larger amount of data.
This especially applies to IoT deployments following the REST architectural style, e.g., those involving the \gls{coap} \cite{CoAPrfc}.
Application scenarios where power-intensive services may be leveraged for DoS and where our solution is applicable (but not limited to) are, e.g., indoor \gls{rtls} \cite{Boulos2012}, livestock monitoring \cite{smart_livestock}, industrial IoT deployments \cite{IIoT} or 
Car2X applications \cite{car2X} where vehicles communicate with constrained sensors embedded in the infrastructure. 
We discuss such use cases in more detail in \Cref{sec:use_case}.

%
Previous work in the area of battery exhaustion attacks leveraging application layer services considers mainly two approaches: 
1) cryptographic puzzles forcing the \Requester{} to perform costly computations to balance the effort of both \Requester{} and \Provider{} \cite{Stajano1999, Hummen2013}  and
2) efficient symmetric authentication of the \Requester{} using pre-shared keys \cite{Brasser2016, Martin2004}.
The main drawback of the former is the adjustment of the puzzle difficulty and verifying that a puzzle solution was calculated by the party for which the puzzle was issued. 
The main drawback of the latter approach is that it does not protect against compromised \Requesters{}. Authentication ensures that the \Requester{} possesses an authentication key, but the key may be leaked or the \Requester{} may be infected with malware using the key.
Further, it is not always possible to establish pre-shared keys, especially when it is not known a priori which devices will communicate with each other.
In such cases, asymmetric authentication algorithms and certificates are commonly used in traditional, unconstrained computer networks. Unfortunately the transfer of certificates in IoT networks with constrained data rates and the verification of asymmetric signatures remains challenging.
In contrast, our approach is to leverage a trusted third party, henceforth called \Backend{}, which has two main responsibilities.
First, to execute an attack detection algorithm on per \Requester{} basis and if a given \Requester{} attempts a battery exhaustion attack to drop its requests and,
second, to offload the computation and communication effort an asymmetric authentication protocol may have from the \Provider{}.
This way we ensure the availability of the services for benign \Requesters{} even in the presence of malicious \Requesters{} and even if the attacker
is capable of spoofing MAC and IP addresses. 
We make the following contributions:
\begin{description}
\item[Contribution 1.]
We are the first to propose the usage of rate limitation algorithms for detecting and throttling battery exhausting attacks.
More precisely, we demonstrate the parametrization of two such alternative algorithms, Leaky Bucket and \gls{ewma} \cite{Rathgeb1991}.
We argue that these algorithms are especially effective as they not only consider common flooding DoS attacks characterized by extreme request rates, but also consider a distinguishing property of battery exhaustion attacks: being feasible by requesting services at inconspicuous rates during less active operational times.

\item[Contribution 2.]
We introduce and formally verify two authentication protocols suitable for different IoT topologies:
\begin{enumerate}
 \item \textit{Backend as a Proxy}: Only the \Backend{} directly communicates with the \Provider{}, but not the \Requesters{}.
 The \Provider{} only returns service responses to \Requesters{}.
 \item \textit{Backend as Ticket Issuer}: The \Backend{} issues \Requesters{} single-use cryptographic tickets which they can use for direct communication with the \Provider{}.
\end{enumerate}
Both protocols ensure service availability for benign \Requesters{} even if \Providers{} are under a battery exhaustion attack.
The protocols achieve 
that the \Provider{} is involved only in lightweight operations whereas
the attack detection algorithm is executed on per \Requester{} basis on the \Backend{}.
For the formal verification we use the state of the art cryptographic protocol verifier ProVerif \cite{proverif}.

\item[Contribution 3.]
We evaluate the Leaky Bucket and \gls{ewma} algorithms for detecting and throttling battery exhaustion attacks through simulation. We also provide a practical implementation of the proposed protocols considering state of the art battery-powered constrained \Providers{} using IPv6 over \gls{ble} communication \cite{rfc7668}. Moreover, our evaluation is centered around a real world \gls{rtls} use case in hospitals, see \Cref{sec:use_case}. As platform for the constrained \Provider{}, we leveraged the widely used nRF51422 \gls{ble} SoC from Nordic Semiconductor. 
\end{description}

\section{Use Cases and Design Goals}\label{sec:use_case}
We present four representative IoT use cases where power intensive services may be leveraged for battery exhaustion attacks. Then, we discuss the use cases' characteristics based on which we derive design goals for our detection algorithm and authentication protocols.

\paragraph{\gls{rtls} in Hospitals}
The usage of \gls{rtls} in hospitals for locating assets such as equipment, supplies and specimens is broadly adopted \cite{Boulos2012, zebra2018, RTLS_centrak}. 
In such systems, the assets are equipped with low-cost tags powered by coin cell batteries and communicate with gateways with known location. The approximate position of a tag can be retrieved by staff members using  interface devices, e.g., tablets or smartphones. 
In order to ease the proximity search for which the accuracy of the \gls{rtls} system is not sufficient a \textit{pick-by-light} \cite{Guo2014} approach is often adopted. For that approach the tags are equipped with an LED which is turned on when the tag's position is requested.
In such settings, a battery exhaustion attack can be conducted by a malicious individual who may capture an interface device and then excessively request position information for some or even all supplies.  
A closely related scenario is the usage of smart tools in industrial IoT settings \cite{hilti}.

\paragraph{Smart Farming}
One smart farming application is to equip livestock with implantable or wearable battery-powered devices capable of monitoring vital parameters and position \cite{smart_livestock}. In this use case, a power intensive service may be retrieving the position of a given animal which requires the activation of a GPS module. An attack can be conducted similar to the previous scenario where adversaries can use interface devices used for retrieving information from livestock. 

\paragraph{Car2X}
Car2X applications may require vehicles to communicate with constrained sensors embedded in roads, bridges or buildings, e.g., for acquiring information about pedestrians around a corner \cite{car2X}. In this example, the power intensive service may be the sensor data acquisition and processing. Attackers may capture the vehicle's ECU responsible for sending the requests and extract the authentication key. Then, the attacker may use this key for excessively requesting services from the transport infrastructure.

\paragraph{Remote Attestation}
Another use case emerges from the active research area of remote attestation for constrained devices. Here, a microcontroller calculates a signed hash of its binary to prove its software integrity upon request from a remote party. Since such computations may be time and resource consuming, excessive attestation requests may be used for battery exhaustion attacks.
This is a well-recognized problem pointed out in several research papers \cite{Brasser2016, Hristozov2018,Ibrahim2017}.

\paragraph{Common Characteristics and Design Goals}
Considering the representative examples we point out several key characteristics based on which we derive design goals DG-I to DG-VI.
\begin{itemize} 
    \setlength\itemsep{1em}
    \item
    The \Provider{} is a battery-powered constrained device communicating over a constrained network.\\
    \textbf{DG-I:}
    The \Provider{} should be involved only in lightweight operations and minimal data transfers.
    \item  
    Providing a service non-interactively, e.g., at a time scheduled by the \Provider{}, is often a suboptimal solution as the service may not be needed at the time it is provided and therefore will be a waste of energy, see for instance the smart farming or remote attestation examples.\\
    \textbf{DG-II:}
    Services should be delivered only at request.
    \item 
    In case of a battery exhaustion attack, it is not adequate to simply deny services to all \Requesters{} to protect the battery's lifetime. Unavailability of services on global scale may have negative consequences in any use case. While unavailability might disturb regular operational processes such as in smart farming, it may also adversely affect the health of humans in extreme cases as applicable in the \gls{rtls} example.\\
    \textbf{DG-III:}
    Attack detection needs to be applicable on per \Requester{} basis instead of on system-wide basis.
    This ensures that those Requesters{} attempting a battery exhaustion attack can be excluded while benign \Requesters{} may still access services.
    Using a global leaky bucket, for example, would result in the denial of services for all \Requesters{} when detecting a battery exhaustion attempt.
    Thus, all requests should be mappable to \Requesters{}. Further, a suitable method must be robust against attackers capable of conducting spoofing attacks. .
    \item  
    In complex deployments like in Car2X or smart tools use cases, it is not practical to establish pre-shared keys between \Requesters{} and \Providers{} for lightweight authentication, i.e, for mapping requests to \Requesters{}.\\
    \textbf{DG-IV:}
    The mapping of requests to \Requesters{} should be possible without requiring pre-shared keys between \Requester{} and \Provider{}.
    \item 
    During the lifetime of the \Provider{}, some \Requesters{} may use a given service more often than others, e.g., a car which often passes by a given senor will request its data a lot more often than cars passing by only rarely, see the Car2X example.
    Therefore, reserving a fixed amount of energy for each \Requester{} is a suboptimal solution requiring bigger batteries.\\
    \textbf{DG-V:}
    The detection and throttling mechanism should not introduce a maximum limit of service requests for each \Requester{} but rather constrain each \Requesters{}' maximum request rate.
    \item
    In some deployments, e.g., those using remote attestation or the Car2X case, the number of \Requesters{} and/or \Providers{} may be extremely high.\\
    \textbf{DG-VI:}
    The detection algorithm should be efficient in terms of memory and computation. This design goal ensures feasible hardware requirements for the \Backend{} and therefore makes our solution scalable for large deployments.
\end{itemize} 

\section{Adversarial Model}\label{sec:adversaries}
We consider an attacker in the network capable of eavesdropping legit communication and sending arbitrary messages to \Backend{}, \Requesters{} and \Provider{}.
This means that the attacker is able to conduct spoofing attacks, i.e., the attacker may fake any kind of public identifiers such as IP and MAC addresses.
We also assume the attacker to be able to perform hard- and software attacks on (a limited amount of) legit \Requesters{} to extract or use their authentication keys before executing the battery exhaustion attack.

We assume that the attacker is unable to conduct other \gls{dos}-causing attacks than the battery exhaustion attack.
First, this means that we assume that the attacker can neither compromise the \Backend{} nor the \Provider{}.
Second, this means the attacker cannot saturate the communication channel of the \Provider{} or any of the \Backend{}'s resources.
Further, the attacker cannot drop or alter communication packets exchanged between legit parties.
This differentiates our attacker in the network from the Dolev-Yao \cite{Dolev1981} model.
Otherwise, the attacker may simply drop the requests of legit \Requesters{} or corrupt the packets and cause \gls{dos} this way.
We also assume that the attacker is unable to break state-of-the art cryptographic primitives.

\section{Detecting Battery Exhaustion Attacks}\label{sec:detection}
Compared to other flooding attacks targeting, e.g., the victim's communication bandwidth, memory or computational capabilities, the battery exhaustion attack is fundamentally different and therefore requires different detection techniques. 
In a typical flooding attack the resources are exhausted when a high number of requests occurs in a short period of time and therefore the detection techniques are optimized to detect such peak request rates. 
Causing battery exhaustion however, can be more subtle. It is sufficient to lightly increase the request rate during usually less active operational times. 

Based on that insight we propose to detect battery exhaustion attacks by averaging the energy drained through requests over a sufficiently long time window and by comparing the average with a predefined threshold depletion rate $\lambda_{th} = E_{tot}/(T\times N)$.
Here, $E_{tot}$ is the device's total available energy for service requests, $T$ is the desired life span of the battery and 
$N$ is the estimated amount of active legit \Requesters{} in a given short period of time.
For calculating the average of the drained energy several techniques exist, for instance
1) Leaky Bucket Algorithm,
2) \gls{ewma}, 
3) the Moving Window Mechanism or the
4) Jumping Window Mechanism
\cite{Rathgeb1991}.     

Compared to the Leaky Bucket Algorithm and \gls{ewma}, the Moving Window Mechanism requires to save the history of past requests in a buffer of the size of the averaging window. Therefore this technique requires significantly more memory resources and may thus be disadvantageous for large deployments, see \Cref{sec:use_case} DG-VI.
The Jumping Window Mechanism may be implemented as a counter which indicates the spent energy. The counter is compared to a fixed threshold to detect an attack.
The threshold takes a burst tolerance margin into account which allows that regular request bursts do not cause false positives. 
The counter is reset in a fixed time interval referred to as jumping window. 
An attacker may exhaust the margin once per window.
In contrast, in the Leaky Bucket Algorithm and \gls{ewma} an attacker may exhaust the burst tolerance margin only once in a \Provider{}'s life-time.
Further, the Leaky Bucket and \gls{ewma} algorithms require merely a single small state variable per \Requester{}, making them suitable for large IoT deployments.
For these reasons, we consider the Leaky Bucket and \gls{ewma} algorithms most suitable in terms of scalability and effectiveness.
We discuss both in detail in \Cref{sec:lb} and \Cref{sec:ewma} and demonstrate how they can be leveraged for detecting battery exhaustion attacks. 
See \cite{Rathgeb1991} for a more detailed discussion of the tradeoffs of those techniques.
To execute the algorithms on a per \Requester{} basis, we authenticate each \Requester{} before executing the rate limitation algorithm, see \Cref{sec:trust_authority}.

\subsection{Leaky Bucket Algorithm}\label{sec:lb}
The Leaky Bucket algorithm \cite{Rathgeb1991} is expressed through the equation:
\begin{equation} \label{eq:LB}
        e_{lb}[n] =   
        \begin{cases*}
          e_{lb}[n-1] + E_s - D      & \small{if request and} $e_{lb}[n]> 0$ \\
          e_{lb}[n-1] - D            & \small{if no request and} $e_{lb}[n] > 0$ \\
          0                     & \small{else.}
        \end{cases*} 
\end{equation}
$e_{lb}$ is a counter incremented at each service request by the amount of energy $E_s$ required to serve the request. $e_{lb}$ is decremented by a fixed predefined amount of energy $D$ per time unit $n$ as long as $e_{lb}$ is positive. The initial state of the leaky bucket is $e_{lb}[n=0] = 0$.
If the rate with which the requests arrive cannot be compensated by $D$ then $e_{lb}$ starts to increase. An attack is detected if
    $e_{lb}[n] > K_{lb}$,
where $K_{lb}$ determines the burst tolerance of the method.
For using the Leaky Bucket algorithm for detecting battery exhaustion attacks the energy requirements $\{E_{s1}, E_{s2}, E_{s3} \ldots \}$ of all services hosted on the \Provider{}, $D$ and $K_{lb}$ have to be defined. 
The energy requirements of the hosted services can be measured during the development of the \Provider{}. 
The decrement $D$ can be set as $D= \lambda_{th}t$, where $t$ is the duration of one time unit.
For calculating $K_{lb}$, the most power intensive request burst to be still considered non-excessive has to be defined and be fed to \autoref{eq:LB}, e.g., ten requests of a given service with energy consumption $E_{s1}$ in ten subsequent time units. The resulting $e_{lb}$ value represents the required depth $K_{lb}$ of the leaky bucket.   

\subsection{EWMA} \label{sec:ewma}
\gls{ewma} \cite{Rathgeb1991} is expressed through the equation:
\begin{equation} \label{eq:EWMA}
   e_{ewma}[n] =   
    \begin{cases*}
        (1-d)E_s +de_{ewma}[n-1]       & \small{if request}  \\
        de_{ewma}[n-1]               & \small{if no request}  \\
        e_0                   & \small{if} $n=0$.
    \end{cases*} 
\end{equation}
Again, $E_s$ is the energy consumption of the requested service. The variable $d= e^{-1/T}$ is a decay parameter in the range $0<d<1$.
$T$ is the desired life span of the battery and $e_0$ is the initial state.
An attack is detected if 
   $e_{ewma}[n] > K_{ewma}$,
where $K_{ewma}$ is a burst tolerance parameter.
In order to use \gls{ewma} for detecting battery exhaustion attacks, the energy requirements of the hosted services $\{E_{s1}, E_{s2}, E_{s3} \ldots \}$, $K_{ewma}$ and $s_0$ have to be defined. Again, the energy requirements of the services can be measured during the design time of the \Providers{}. $K_{ewma}$ can be calculated similar to $K_{lb}$ for the Leaky Bucket algorithm. The initial state can be set as $e_0 = \lambda_{th}t$, where $t$ is one time unit. 

We evaluate the usage of both the Leaky Bucket algorithm and EWMA for detecting malicious requesters through simulation of a realistic IoT deployment in our evaluation in \Cref{sec:evaluation}.

\FloatBarrier 

\section{Authentication Protocols} \label{sec:trust_authority}
\begin{figure}[t]
\centering
\begin{subfigure}[c]{0.49\textwidth}
\includegraphics[width=\textwidth]{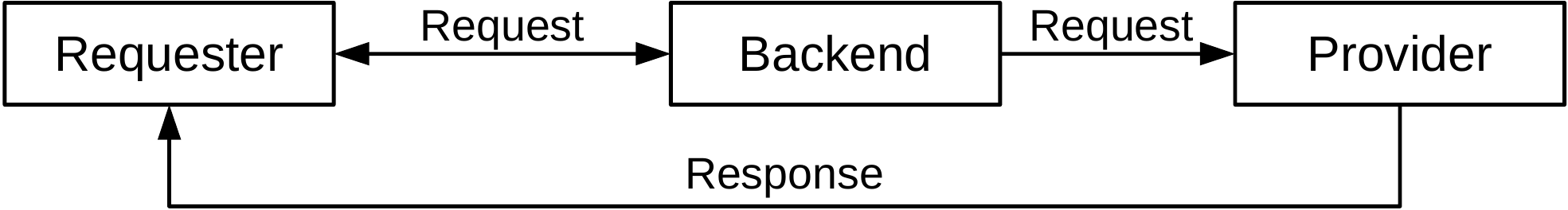}
\subcaption{Backend as a Proxy}
\label{fig:secv_intro_proxy}
\end{subfigure}
\begin{subfigure}[c]{0.49\textwidth}
\includegraphics[width=\textwidth]{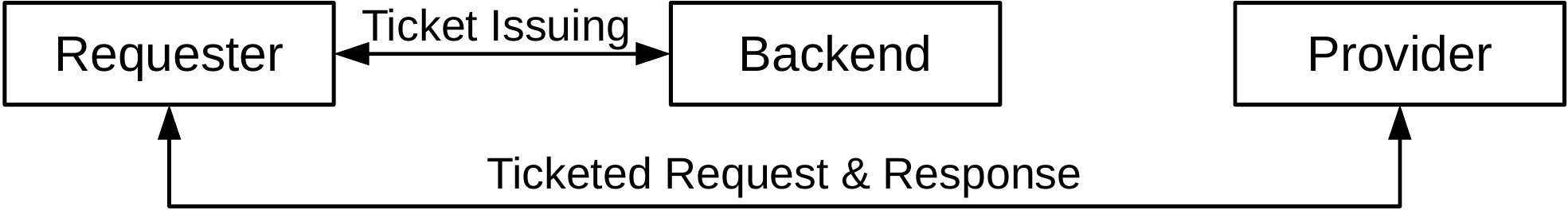}
\subcaption{Backend as Ticket Issuer}
\label{fig:secv_intro_tickets}
\end{subfigure}
\caption{Overview of the authentication protocols.}
\label{fig:secv_intro}
\end{figure}

In this section, we propose two authentication protocols suitable for different IoT communication models and describe their formal verification using ProVerif \cite{proverif}.
Both protocols leverage a trusted \Backend{} as an intermediary between \Requester{} and \Provider{}, as depicted in \autoref{fig:secv_intro}.
We place the \Requester{} authentication and attack detection algorithm logic in the \Backend{} to keep the complexity on \Provider{}-side minimal.
The protocol \textit{Backend as a Proxy} allows direct communication from \Backend{} to \Provider{}, but not from \Requester{} to \Provider{}.
The protocol \textit{Backend as Ticket Issuer} allows direct communication between \Requester{} and \Provider{} when the \Requester{} possesses a valid cryptographic ticket issued by the \Backend{}.
In both protocols, the \Requester{} authenticates to the \Backend{} using asymmetric cryptography while the \Backend{} authenticates to the \Provider{} using lightweight symmetric cryptography.
The required symmetric keys for \Backend{} to \Provider{} authentication can be provisioned in a secure environment before the enrollment, such as in the production facility at the time the firmware is flashed onto the \Provider{}.
This way we achieve 1) that the attack detection algorithm can be executed on per \Requester{} basis on the \Backend{}, therefore allowing to drop malicious requests while sustaining availability for benign \Requesters{} and 2) that the \Provider{} is involved only in lightweight operations with low communication overhead.

In general, we assume that the \Requester{} is a more powerful device capable of communicating with the \Backend{} through unconstrained networks, e.g., the Internet. We assume that the constrained battery-powered \Provider{} communicates through a constrained network.
The two protocols we propose are suitable for the following two IoT network topologies.
First, the \Backend{} may directly communicate with the \Provider{} via an edge gateway over the Internet.
Such topology can, e.g., be present in the RTLS use case from \Cref{sec:use_case}.
Second, the \Requester{} may directly communicate with the \Provider{} through a constrained networks, e.g., in the Car2X case.
We design the protocol \textit{\Backend{} as a Proxy} (see \Cref{sec:baap}) for the former case and protocol \textit{\Backend{} as Ticket Issuer} (see \Cref{sec:bati}) for the latter.
Both authentication protocols require:
\begin{enumerate}
	\item The \Requester{} and \Backend{} to have asymmetric key pairs and certificates along with trust anchors for certificate chain verification.
	\item The \Provider{} and \Backend{} to share a symmetric key provisioned, e.g., during the \Provider{}s' production.
	\item The \Provider{} to use a lower layer protocol supporting reliable message transport, e.g., CoAP or TCP, where messages are acknowledged and retransmitted at loss.
\end{enumerate} 
%

We introduce the following notation for the protocols. The reader may skip this part and refer to it while reading the protocols.
\begin{protocolbox}
\textbf{Notation.}
\begin{enumerate}\itemsep0em
	\item $ID_S$, $ID_R$, $ID_P$ are identifiers of the requested service, the Requester and the \Provider{}, respectively.
	\item $N$ is a nonce.
	\item $i$ is an anti-replay counter. 
	\item $F$ is a flag indicating the receiver whether it has to send its certificate in the subsequent message. If the party sending the flag can cache certificates, the flag can be set only at the very first protocol run.
	\item ${C_X}^*$ is the certificate of party X. The asterisk denotes that the certificate may not always be transmitted, depending on $F$.
	\item $S_X(m)$ is an asymmetric signature of the message $m$ generated with the private key of party X.
	\item $K_X^{pub}$ is the public key of party X.
	\item $K_{XY}$ is a shared secret key between party X and Y.
	\item $M_{K_{XY}}(m)$ is a Message authentication Code (MAC) over $m$ with key $K_{XY}$ shared between party X and Y. 
	\item $r$ is a random number and $h$ is the hash of $r$ calculated with a preimage resistant hash function.
	\item $T$ is a cryptographic ticket allowing a single request of a service.
\end{enumerate}
\end{protocolbox}

\subsection{\Backend{} as a Proxy}\label{sec:baap}
In the following, we describe a method where the \Backend{} serves as proxy between \Requester{} and a \Provider{}, see \autoref{fig:secv_intro_proxy}. When the \Backend{} receives a request, it executes one of the detection algorithms from \Cref{sec:detection} and if the current request is considered benign it is forwarded to the given \Provider{}, as detailed in Protocol \ref{protocol:proxy}.
\begin{protocol}{Backend as a Proxy}\label{protocol:proxy}
		\item \Requester{} $\rightarrow$ \Backend{}: $ID_R$, $N_1$, $F$
		\item \Requester{} $\leftarrow$ \Backend{}: $N_2$, $F$, ${C_B}^*$, $S_B(N_1, N_2)$
		\item \Requester{} $\rightarrow$ \Backend{}: $ID_S$, $ID_P$, $S_R(ID_S,ID_P,N_2)$, ${C_R}^*$
		\item \Backend{} $\rightarrow$ Provider: $ID_S$, $M_{K_{PB}}(ID_S,ID_P,i)$	
\end{protocol}
%
A new protocol run is initiated by the \Requester{} by sending its identifier $ID_R$, a nonce $N_1$ and a certificate flag $F$ to the \Backend{} in message (a). The certificate flag is set if the \Backend's certificate is not in the \Requester{}'s local cache.
%
In message (b) the \Backend{} sends a new nonce $N_2$, a certificate flag $F$ depending on whether it possesses a certificate corresponding to the received $ID_R$ and optionally its certificate $C_B$, depending on the flag received in message (a). 
In addition, the signature $S_B()$ over $N_1$ and $N_2$ is generated and sent.
When the \Requester{} receives message (b), it verifies $S_B()$ and $C_B$ in case it was requested with the flag $F$ in message (a).
The signature $S_B()$ and the nonces $N_1$ and $N_2$ prevent attackers from forging messages (b).
%
In message (c), the \Requester{} generates and sends a signature over $ID_S$, $ID_P$ and $N_2$. The \Requester{} sends also $ID_S$ and $ID_P$ in plaintext. Depending on the flag received in message (b), the certificate of the \Requester{} $C_R$ is also sent. 
The replay protection of message (c) is achieved through the the signature and nonce $N_2$. 
Receiving the message (c), the \Backend{} checks the signature and the nonce, executes the detection algorithm and sends message (d) if the current request is considered non-excessive. 
When the \Provider{} receives message (d), it checks the lightweight MAC and the replay protection counter $i$ before serving the request.
We omitted the subsequent procedure of serving the request from our protocols as this step is highly specific to the use case.

\subsection{\Backend{} as Ticket Issuer}\label{sec:bati} 
In this protocol, the \Backend{} issues the \Requester{} single-use permission tickets allowing to directly request services from the \Provider{}, see \autoref{fig:secv_intro_tickets}.
\begin{protocol}{\Backend{} as Ticket Issuer}\label{protocol:tickets}
		\item \Requester{} $\rightarrow$ \Backend{}: $ID_R$, $N_1$, $F$
		\item \Requester{} $\leftarrow$ \Backend{}: $N_2$, $F$, ${C_B}^*$, $S_B(N_1,N_2)$
		\item \Requester{} $\rightarrow$ \Backend{}: $ID_P$, $ID_S$, $N_3$, $h$, ${C_R}^*$, $S_R(ID_P, ID_S, N_2, N_3,h)$ 
		\item \Requester{} $\leftarrow$ \Backend{}: $T$, $S_B(T, N_3)$
		\item \Requester{} $\rightarrow$ Provider: $r$, $T$ 
\end{protocol}
Messages (a) and (b) are identical to Protocol \ref{protocol:proxy}. 
Before sending message (c), the \Requester{} chooses a random number $r$ and calculates its hash $h$ using a preimage resistant hash function. Both $r$ and $h$ are used for binding a ticket to a given \Requester{}.
Message (c) contains the identifier of the \Provider{} $ID_P$, the identifier of the service $ID_S$, a new nonce, $N_3$, the hash of the random number $h$, optionally the certificate of the \Requester{} $C_R$ and a signature $S_R(ID_P, ID_S, N_2, N_3,h)$. 
The replay protection of message (c) is accomplished through the nonce $N_2$ and the signature.

When the \Backend{}  receives message (c) it verifies the signature $S_R()$ and, if requested, the certificate $C_R$. After that, the \Backend{} executes the detection algorithm. If the current request is considered non-excessive a ticket $T \coloneqq ID_S, i, M_{K_{PB}}(ID_P, ID_S, h, i)$ is created and sent together with a signature over it and the nonce $N_3$ to the \Requester{} in message (d).
In order to prevent attackers using an intercepted ticket contained in message (d) for requesting a service, the MAC of the ticket contains the hash $h$ of the random number $r$, which is only known to the legit \Requester{}. 
In message (e), the \Requester{} forwards the ticket $T$ to the \Provider{} together with the random number $r$.
When the Provider receives the ticket $T$ it calculates $h$  and verifies the MAC $M_{K_{PB}}()$.
Doing so, we ensure that an attacker cannot use a ticket that has been  eavesdropped in message (d). 

Still, the attacker may intercept a ticket and a random number $r$ contained in message (e) and try to replay them for requesting a service.
We use the counter $i$ contained in the ticket to prevent such replay attacks.
To ensure that replay protection also works reliably when the \Provider{} receives valid tickets in different order than issued by the \Backend{} we leverage a small memory cache.
This cache is initially empty and when a ticket with a valid MAC is received we save its counter in the cache.
For each ticket received with a valid MAC, we check
1) if its counter is in the cache and
2) if the counter is more distant than a validity distance $\Delta i$ from the highest counter contained in the cache. If any of these conditions applies we discard the request. Otherwise, we save the current counter in the cache and provide the service.
When saving a new counter in the cache we check if this is the new highest counter value.
If so, we remove all counters more distant than $\Delta i$ from the cache. 

\FloatBarrier 
\subsection{Formal Verification with ProVerif}
In this section we describe our verification of our proposed protocols using ProVerif \cite{proverif}. For both protocols we developed ProVerif verification models which we made available online \cite{proverifModels}.

ProVerif is a command line tool which expects a model of a cryptographic protocol and a set of security properties to be verified as input, both encoded in the language of typed pi calculus. 
The protocol properties relevant in our scope and which ProVerif can prove are:
1) reachability -- an attacker cannot deduce a certain value, e.g., a secret key, and
2) correspondence assertions -- a certain protocol state can only be reached if the protocol was in a certain other state before \cite{Blanchet2018}.

In ProVerif, cryptographic primitives are abstracted and assumed to be perfect.
They are modeled through constructors and destructors. Constructors and destructors only model the relations between input and output parameters but not the mathematic operations which a cryptographic primitive conducts. A constructor models a forward cryptographic primitive, e.g., the construction of an asymmetric signature and a destructor the reverse primitive, e.g., verification of an asymmetric signature. When no reverse primitive exists, e.g., a hash function, the primitive is modeled only as a constructor.

In a ProVerif model, the protocol participants are modeled as parallel processes communicating through a public channel. 
The attacker has no knowledge about the operations inside the processes but has full control over the messages exchanged on the public channel. 
ProVerif considers Dolev-Yao \cite{Dolev1981} attackers capable of eavesdropping, deconstructing and dropping legit messages, constructing and injecting new ones and using known crypto primitives. However, the attacker cannot break crypto primitives, e.g., construct a MAC without the key. 
Comparing the capabilities of our attacker in \Cref{sec:adversaries} related to attacks on the communication channel with the Dolev-Yao model, our attacker is slightly less powerful.
The difference is that we assume that our attacker cannot drop or alter legit messages, which would otherwise be sufficient for conducting DoS. Therefore, if protocols \ref{protocol:proxy} and \ref{protocol:tickets} are correct under a Dolev-Yao attacker, the protocols are also correct under the attacker described in \Cref{sec:adversaries}.

For modeling the \gls{pki} we used an abstraction where a certificate contains a set of participant identifiers, its public key, and a signature generated with the private key of the \gls{ca}. We modeled the \gls{ca} as a public-private key pair in the main process.

In protocols \ref{protocol:proxy} and \ref{protocol:tickets} we used counters for replay protection.
Because of its internal abstraction
ProVerif is incapable of fully handling protocols involving counters. This is a known limitation for which the authors of ProVerif demonstrate a semiautomatic workaround in \cite{Blanchet2017} which we also leveraged for our models.
In this workaround the counter value is received by the honest participants on the public channel from the attacker, therefore this value may be repeated.
Then, using ProVerif it can be verified that a protocol message which contains a counter value was sent by an honest participant before it was received and processed by another honest participant. This may happen eventually multiple times with the same counter value if the attacker inputs always the same value.
Therefore, in the next analytical step we constrained the attacker to only be able to send counter values in increasing order.
This means that a message containing a given counter value can be received exactly once after it was sent.

To conclude, our models \cite{proverifModels} demonstrate that protocols \ref{protocol:proxy} and \ref{protocol:tickets} have the following security properties:
\begin{enumerate}
	\item No secret values are leaked. For this property, we leveraged the ProVerif capabilities to prove reachability.
    Thus, the secret keys of the \gls{ca}, \Requester{} and \Backend{} and the symmetric key $K_{PB}$ cannot be deduced by the attacker.
	\item A service can only be requested by a \Requester{} possessing a private key.
    For this property, we leveraged the ProVerif capabilities to prove correspondence assertions and the workaround as described above.
\end{enumerate}

\section{Discussion}

In the following discussion, we refer back to the design goals we initially set and discuss a \gls{ddos} attack scenario in which the effectiveness of our protection may be limited.

\paragraph{Design Goals}
Both protocols fulfil our design goals.
In both, the \Provider{} is involved only in lightweight operations, either processing tickets or directly verifying the authenticity the \Backend{} (DG-I) for providing services only at request (DG-II).
The authentication of the \Requester{} to the \Backend{} based on public-key cryptography in combination with per-\Requester{} rate limitation ensures DG-III and does not require preshared keys between each \Requester{} and \Provider{} (only between \Backend{} and \Provider{}), as defined in DG-IV.
Using the Leaky Bucket Algorithm or EWMA for rate limitation, our method does not limit the total amount of possible requests for each \Requester{}, but regulates the rate with which requests are served (DG-V). Moreover, both algorithms are very efficient in terms of computation and memory, thus our solution is scalable for use cases with large amounts of \Requesters{} and/or \Providers{} (DG-VI).

\paragraph{Distributed Denial of Service}
A theoretically possible scenario may be \gls{ddos} where the attacker compromises a large group of \Requesters{} to quickly deplete a Provider's battery by sending requests from each compromised Requester's below its threshold rate. This attack is limited by the following factors:
\begin{enumerate}
\item Each \Requester{} has a unique and independent authentication key. The attacker must thus compromise a large number of \Requesters{} for the attack to scale. Each \Requester{} may be a different device type (e.g. a smartphone or tablet) with different software stacks and thus not expose a single, common vulnerability.
\item The energy each \Requester{} can draw is limited to $\lambda_{th}t$.
\end{enumerate}

A logical approach to protect the battery capacity in the \gls{ddos} scenario could be to use a single rate limiter executed locally on the \Provider{} without \Requester{} authentication. This approach falls short in ensuring availability of services to benign \Requesters{} (DG-III) since a single \Requester{} may cause all incoming requests to be dropped by saturating the rate limiter. Therefore, this approach only shifts the battery exhaustion vulnerability to a \gls{dos} vulnerability feasible by saturating the single rate limiter.
 
\section{Evaluation}\label{sec:evaluation}
We evaluate the proposed detection algorithms and authentication protocols considering the use case \emph{RTLS in hospitals} we previously introduced in \Cref{sec:use_case}. For that purpose we 
1) demonstrate the operation of the Leaky Bucket and EWMA algorithms through MATLAB simulation and
2) implement a proof of concept setup based on the nRF51422 BLE SoC which we use for evaluating the protocols in terms of power consumption and latency. 

\subsection{Simulation}
We first describe how we selected the parameters for our simulation.
Then we demonstrate the severity of the battery exhaustion attack by chaining regular burst requests. We simulate different burst intensities and show their effect respective to the attack duration. In \Cref{sec:simulation} we demonstrate the operation of the Leaky Bucket and EWMA algorithms.

\subsubsection{Simulation Parameters}\label{sec:parametrization}
In the RTLS use case the \Provider{} is a localization tag, see \Cref{sec:use_case}. For the simulation we assumed that it is powered by a CR2430 coin cell battery with capacity of $E_{bat} = 840\,\mathrm{mWh} = \mathrm{3,024\,J}$ \cite{vartaCR2430}. 
We assumed that it hosts a single energy expensive service which flashes an indicator LED requiring $E_s=45\,\mathrm{mJ}$.
We specify $T=1\,\mathrm{year}$ as the operational time of the \Provider{}.
We assume that, in average, $N = 100$ \Requesters{}, i.e., staff members, are active. 

In order to achieve the given operational time using the limited energy budget provided by the coin cell battery the \Provider{} has to keep its hardware modules turned off as long as possible. 
Since the RX part of the radio transceiver is one of the main power dissipators (e.g. 9.7\,mA for nRF51422 \cite{nRF51422}), the capability to receive data, i.e., the latency for receiving a request, and minimizing power consumption are two contradictory requirements.
In our setup we configured the BLE radio to be capable of receiving 20\,bytes of payload data every single  second. Doing so we measured an average current consumption $\overline{i_{rx}}=24\,\mu\mathrm{A}$  on the nRF51422 device. Using $\overline{i_{rx}}$ we calculate the energy $E_{rx}=u\overline{i_{rx}}T= 2,270 \,\mathrm{J}$ required for receiving data throughout the lifetime $T$ with voltage $u=3\,\mathrm{V}$. Using $E_{bat}$, $E_{rx}$ and assuming that the device may require some additional energy for other purposes which we chose to be 10\% of $E_{bat}$ for the given evaluation, we calculate $E_{tot} = E_{bat}-E_{rx} - 0.1E_{bat} = 425\,\mathrm{J}$ and $\lambda_{th} =  12.38\,\mathrm{mJ/day}$. 

Furthermore, we assume that each of the 100 legit \Requesters{} requests services randomly with a probability of request occurrence per day $P=N/T=0.2740$.  
For the evaluation of the Leaky Bucket and \gls{ewma} algorithms in \Cref{sec:simulation} we set their parameters as: 
$D=e_0= 12.38\,\mathrm{mJ}$,  
$K_{lb}=0.4049\,\mathrm{J}$ and $K_{ewma}=\num{9.332e-6}\,\mathrm{J}$.
Note that we calculated $K_{lb}$ and $K_{ewma}$ by evaluating \autoref{eq:LB} and \autoref{eq:EWMA}, respectively, assuming that the most energy consuming but still tolerated burst consists of 10 requests in 10 minutes.
All parameters are summarized in \autoref{tab:setupParameters}.
\begin{table}[t]
	\footnotesize
	\centering
	\begin{tabular}{l@{\hskip 1em}l@{\hskip 1em}l}
		\toprule
		Parameter                             & Symbol              & Value          \\\midrule
		Battery capacity                      & $E_{bat}$           & 3,024\,J        \\ 
		Energy for non-significant purposes   & $E_{else}$          & 10\% $E_{bat}$ \\ 
		Energy cost of service                & $E_{s}$             & 45\,mJ         \\ 
		Desired lifetime                     & $T$                 & 365 days       \\ 
		Active \Requesters{}  in a given time & $N$                 & 100            \\ 
		Average RX current 			         & $\overline{i_{rx}}$ & 24\,$\mu$A     \\ 
		Energy spent for RX in lifetime $T$  & $E_{rx}$            & 2,270\,J        \\ 
		Total energy for requestable services & $E_{tot}$            & 452\,J        \\ 
		Threshold depletion rate              & $\lambda_{th}$      & 12.38\,mJ/day  \\  
		Leaky bucket decrement, \gls{ewma} initial state             	& $D$, $e_0$      & 12.38\,mJ \\  
		Leaky bucket detection threshold   	& $K_{lb}$      & 0.4049\,J \\  
		\gls{ewma} detection threshold   	& $K_{ewma}$      & \num{9.332e-6}\,J \\  
		\bottomrule
	\end{tabular}
	\caption{Summary of the setup parameters for the evaluated IoT use case.}
	\label{tab:setupParameters}
\end{table}

\subsubsection{Attack Severity}
\begin{figure}[t] 
	\centering  
	\includegraphics{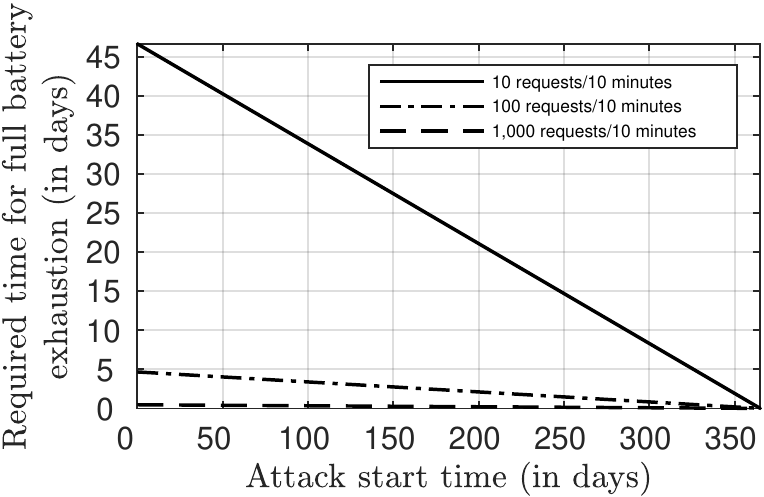} 
	\caption{Time required for full battery exhaustion by chaining regular bursts with different intensity.}
	\label{fig:attack_duration} 
\end{figure}

In \autoref{fig:attack_duration} we plot the time required for a single \Requester{} to cause complete battery exhaustion by chaining regular bursts when no protection is in place and when considering the parametrization as given in \autoref{tab:setupParameters}. From \autoref{fig:attack_duration} it is evident that the required time for a complete battery exhaustion depends on the start time of the attack, i.e., the amount of energy still available in the battery and the proportion of burst energy and burst duration of the most power intensive regular burst. 
If the most intensive burst consists of 10 requests in 10 minutes the attacker needs around 45 days to completely exhaust the battery but if the the most intensive burst consist of 1,000 requests in 10 minutes, the attacker requires less than a day.

\subsubsection{Attack Detection and Throttling through Rate Limitation}\label{sec:simulation}
\begin{figure*}[t] 
	\centering  
	\includegraphics{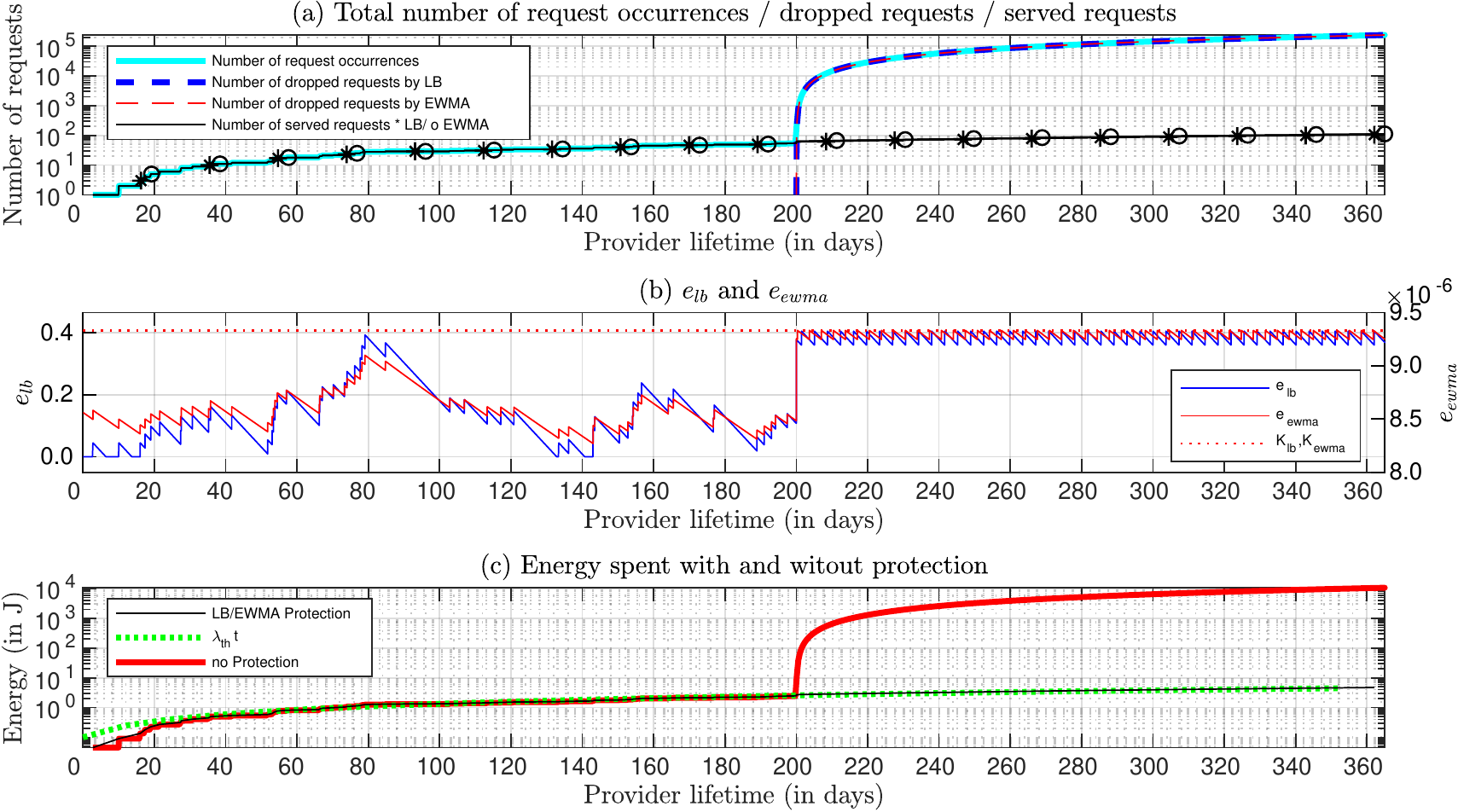} 
	\caption{Attack detection and throttling with Leaky Bucket and \gls{ewma}.}
	\label{fig:lb_ewma} 
\end{figure*}

In order to demonstrate the effectiveness of the rate limitation algorithms for detecting and throttling battery exhaustion attacks we simulated the attack scenario in which a malicious \Requester{} chains regular bursts, see \autoref{fig:lb_ewma}. 
\autoref{fig:lb_ewma}.a shows the number of requests made by the \Requester{} within $T=365$\,days. In the first 200 days the \Requester{} is benign but conducts a battery exhaustion attack after that period (days 200 to 365).
During the legit operational time the \Requester{}
uses the service with probability $P$ for a request.
In \autoref{fig:lb_ewma}.a  we see that the number of requests dropped by the Leaky Bucket and EWMA algorithms rapidly grows during the attack (notice the logarithmic scale of the y axis) and that the number of served requests during the attack is equal for both Leaky Bucket and EWMA algorithms. Therefore, both algorithms are equally suitable for detecting battery exhaustion attacks.   
\autoref{fig:lb_ewma}.b shows the development of $e_{lb}$ and $e_{ewma}$ for the request series from \autoref{fig:lb_ewma}.a. 
In \autoref{fig:lb_ewma}.b we see that as soon as the attack is conducted, $e_{lb}$ and $e_{ewma}$  start to increase until they reach their thresholds $K_{lb}$ and $K_{ewma}$, respectively. When this happens the attack is detected. 
The \Backend{} will allow only such amounts of requests so that the threshold energy resulting from $\lambda_{th}t$ is not exceeded, see \autoref{fig:lb_ewma}.b.
\autoref{fig:lb_ewma}.b also shows the energy a malicious \Requester{} can drain when no protection mechanism is in place (notice the logarithmic scale of the y axis). 

\subsection{Implementation of the Evaluation Setup}\label{sec:implementaion}
\begin{figure}[t] 
	\centering
	\includegraphics[width=\linewidth]{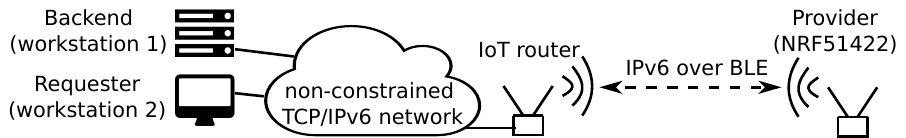}
	\caption{Setup of the IoT network used for the evaluation.}
	\label{fig:attack_overview}
\end{figure}

We sought to realize an abstract IoT scenario with a battery-powered constrained \Provider{}, suitable for a representative evaluation of both protocols.
The following paragraphs elaborate on the most relevant factors that influence the Provider's power consumption and protocol latency.
These are the setup of the IoT network, as well as the protocol-related implementation aspects.
\paragraph{Overview of the Setup}
\autoref{fig:attack_overview} shows the setup of our IoT network.
The setup consists of two Linux workstations, one representing a \Requester{} and the other one representing a \Backend{}.
Further, a \gls{ble} board as a \Provider{} and an IoT border router. We used two separate IPv6 networks, which the IoT border router connects: 
1) a standard non-constrained network for the communication between \Requester{}, \Backend{} and the border router and 
2) a constrained network where the BLE board communicates with the IoT border router using IPv6 over BLE \cite{rfc7668}. 
We use BLE for the constrained network, because BLE can be easily integrated into existing non-constrained IPv6 networks while providing excellent low power characteristics, suitable for battery-powered devices. 
As a representative platform for a constrained \Provider{}, we used the nRF51422 BLE SoC from Nordic Semiconductor. The nRF51422 is based on a Cortex-M0 processor running at 16\,MHz and has 256\,KB of flash memory, 32\,KB RAM and an AES accelerator.

\paragraph{Communication Model and Underlying Protocols}
For each of the two protocols proposed in \Cref{sec:trust_authority}, we used the following communication model and protocols:
\begin{enumerate}
	\item \Backend{} as a Proxy: The \Requester{} is a TCP client while the \Backend{} is a TCP server and at the same time a CoAP client. The \Provider{} is a CoAP server.
	\item \Backend{} as Ticket Issuer: The \Requester{} is a TCP and CoAP client. This \Backend{} is a TCP server and the \Provider{} is a CoAP server.
\end{enumerate}
For the protocol implementations for \Requester{} and \Backend{} we leveraged the standard Linux TCP API and the open source CoAP library \textit{libcoap} \cite{libcoap}. For the BLE SoC, we used the CoAP stack provided by Nordic Semiconductor's SDK \cite{nordicSDK}.

\paragraph{Request Message Format}
The size of the last message in protocols \ref{protocol:proxy} and \ref{protocol:tickets} is of critical importance as it is sent over the constrained network to the Provider and may therefore significantly influence protocol latency, i.e. the time required for requesting a service. In our implementation we used C structures for defining each message, see \autoref{lst:req_msg}. According to the listing, message (d) for protocol \ref{protocol:proxy} requires 9 bytes and message (e) for protocol \ref{protocol:tickets} 15 bytes.
Those messages contain several values which size is either 
1) dependent on the concrete IoT use case ($ID_s$ and $i$) or
2) a tradeoff between size of the complete message and security against guessing attacks ($r$ and $MAC$).  
For our prototype we used a MAC of length 8 bytes which is considered secure by NIST \cite{Dworkin2016}. Using a sufficiently long MAC is essential for our protocols since the adversary can otherwise construct a valid message to
1) be used for requesting a service and/or
2) be used to put the counter out of synchronization.
Another parameter which length is a tradeoff between security against guessing attacks and the size of the message is $r$ in Protocol \ref{protocol:tickets} message (e). The length of $r$ should be chosen depending on a ticket's validity time.
Since a ticket may be significantly shorter living than the pre-shared authentication key, $r$ may be chosen to be shorter than the MAC.

\begin{listing}
    \footnotesize
    \centering
     \begin{sublisting}[b]{0.4\linewidth}
      \begin{lstlisting}[gobble=8]
        typedef struct {
            byte ID_s;
            byte MAC[8]; 
        } prot1_msg_d_t;
      \end{lstlisting}
      \vskip.5cm
      \caption{Protocol \ref{protocol:proxy} msg d}
    \end{sublisting}
 %
    \begin{sublisting}[b]{0.4\linewidth}
        \begin{lstlisting}[gobble=8]
            typedef struct {
                byte r[4];    
                byte ID_s;
                byte i[2];
                byte MAC[8]; 
            } prot2_msg_e_t;
        \end{lstlisting}
        \caption{Protocol \ref{protocol:tickets} msg e}
    \end{sublisting}
    
    \caption{Data structures of the request messages sent to the Provider.} \label{lst:req_msg}

\end{listing}

\paragraph{Hash and MAC Functions}
To prevent a malicious \Requester{} from leveraging the \Provider{}'s MAC and hash computation for battery exhaustion, these primitives must be efficiently computable.
In order to achieve this in our implementation we leveraged the AES hardware accelerator available on \gls{ble} SoC. Note that, such an accelerator is a common hardware peripheral for this class of devices. For the implementation of the MAC and hash functions we used the CBC-MAC and Davies-Meyer \cite{Menezes1996} hash function, respectively.

\paragraph{Alternative Asymmetric Protocol}
The most obvious alternative to our protocols is an asymmetric protocol where the \Requester{} requests services directly from the \Provider{} by sending its certificate and a signature over the service ID, \Provider{} IDs and replay protection counter. 
When the \Provider{} receives a request, it verifies the certificate and signature, executes the attack detection algorithm and provides its service in case of a positive result.
We implemented this asymmetric protocol as baseline for comparison with our protocols from \Cref{sec:trust_authority}.
For the asymmetric protocol we used the ED25519 signature algorithm and X.509 certificates.
For the verification of the signatures on the nRF51822 SoC we used the wolfcrypt library~\cite{wolfcrypt2019}, because the SoC lacks an accelerator for asymmetric signatures.
To the best of our knowledge, there is no such peripheral for low-end \gls{ble} SoCs (e.g. Cortex M0) on the market.
We only found higher-end \gls{ble} SoCs (Cortex M4 and above) equipped with an accelerator for asymmetric cryptography.

\subsection{Request Latency and Cost of Authentication}
In this part, we compare the protocol latencies and cost of authentication 
for protocols \ref{protocol:proxy} and \ref{protocol:tickets} with each other and with our asymmetric baseline protocol.
\begin{table}[t]
	\footnotesize 
	\centering
	\begin{tabular}{l@{\hskip 2em}r@{\hskip 2em}r@{\hskip 2em}r}
		\toprule
		Protocol    & Req. Size (byte) & Latency (s) & Energy Auth. (J) \\ \midrule
		Proxy   	& 9                & 1           & \num{1.21e-6}    \\ 
		Ticket Issuer & 15                     & 1           & \num{2,34e-6}    \\ 
		Asymmetric  & 532                    & 27          & \num{33.14e-3}   \\ \bottomrule 
	\end{tabular}
	\caption{Number of bytes per request message, latency and energy cost for authentication for protocols \ref{protocol:proxy} and \ref{protocol:tickets} in comparison with asymmetric authentication.}
	\label{tab:results}
\end{table}

\begin{table}[t]
	\footnotesize 
	\centering
\begin{tabular}{l@{\hskip 1em}r@{\hskip 1em}r}
	\toprule
	Protocol    & Energy (J) & \% of $E_{bat}$ \\ 
	\midrule
	Proxy   & 38         & 1.14            \\ 
	Ticket Issuer & 74         & 2.23            \\ 
	\bottomrule        
\end{tabular}
\caption{Energy drain by forcing the \Provider{} to authenticate one attacker-injected request per second over its whole lifetime.}\label{tab:symm_req_energy}
\end{table}

We summarized our results in Tables~\ref{tab:results} and \ref{tab:symm_req_energy} based on which we conclude the following:
According to \autoref{tab:results}, asymmetric authentication with certificates causes high latencies in constrained networks (in our case 20 Bytes/s receive data rate) and therefore may be not applicable in many real use cases.
In contrast, our authentication protocols are many times faster and therefore, in this aspect, superior.
Second, using the commonly available AES accelerator on the radio SoC, we are capable of authenticating requests very efficiently in terms of energy. 

In \autoref{tab:symm_req_energy} we present an extrapolation of the measurements summarized in \autoref{tab:results}. Note that we assume a device powered by a coin cell battery with capacity $E_{bat}= 3,024\,\mathrm{J}$ \cite{vartaCR2430}. For that device we assume that its desired lifetime is one year.
\autoref{tab:symm_req_energy} clearly shows that attacks where an attacker injects requests (Protocol 1) or invalid tickets (Protocol 2) to the \Requester{} to force the \Requester{} to spend energy for their authentication does not cause battery exhaustion. An attacker is only capable of exhausting $1.14\%$ of $E_{bat}$ when Protocol 1 is used and $2.23\%$ of $E_{bat}$ when Protocol 2 is used.   


\section{Related Work}\label{sec:related_work}
In the following, we investigate previously published battery exhaustion attacks and their countermeasures. Additional systematization of knowledge in this area can be found in \cite{Merlo2015}.
%

In \cite{Stajano1999,Stajano2002}, the authors describe an attack using legit service requests to prevent the \Provider{} from entering sleep mode, thus exhausting its battery. 
The authors briefly mention the use of cryptographic puzzles as a possible solution without presenting specific methods or evaluations. Cryptographic puzzles \cite{Juels1999, Aura2000, Dean2001, back2002} are mathematical tasks with adjustable complexity determined by the \Provider{}. The puzzles have to be solved by the \Requester{} before the service is provided. Puzzles aim to balance the effort of both parties to slow down the request rate of attackers.
The approach introduces two challenging requirements:
The first one is that the difficulty of the puzzle has to match the computational capabilities of the \Requester{}. Otherwise,
very weak devices will not be capable of solving the puzzle at all and strong devices will quickly solve it, i.e., the puzzle will not serve its purpose to slow down the request rate.  
To satisfy this requirement, information about the computational capabilities of the \Requester{}s has to be communicated to the \Provider{} in a secure and trusted way.
The second requirement is that the \Provider{} needs a mechanism to verify whether the puzzle solution was indeed calculated by the \Requester{}, otherwise a malicious \Requester{} may outsource the computation to more powerful devices.
Because of those challenging requirements, we consider puzzles unsuitable. 

In \cite{Hristozov2018,Brasser2016}, the authors elaborate on the case of runtime attestation of constrained devices. In their work, a constrained device calculates a signed hash of its binaries in order to prove the software state to a remote party. Since this may be a time consuming calculation, excessive attestation requests may be used for a DoS attack. The solution proposed in \cite{Brasser2016} is based on symmetric authentication.
In addition to the difficult symmetric key establishment in many IoT deployments, this approach also has the major downside that it does not prevent compromised \Requesters{} from conducting a battery exhaustion attack.

A body of research in the area of battery exhaustion attacks on mobile computing devices exists as well \cite{Martin2004, Nash2005, Buennemeyer2007a, Moyers2010, Buennemeyer2007b}. The authors in \cite{Martin2004} consider an attack detection and identification method with \Requester{} authentication and power signature monitoring hosted on the service \Provider{}. 
For the authentication, the authors describe an abstract method of \textit{multiple layers of authentication} without focusing on any specific algorithms. To the best of our knowledge, this method cannot avoid the use of intensive cryptographic operations without using pre-shared secrets and may therefore present an additional battery exhaustion attack vector when used for constrained devices.
Moreover, the proposed power signature monitoring requires hardware for power measurement, a database with reference power traces hosted on the service \Provider{} and a comparison algorithm to match the reference traces with the current measurement. Those requirements inevitably lead to increased demands in terms of memory and computational power and are therefore not suitable for constrained devices.
In \cite{Nash2005} the authors consider mobile computers executing several parallel processes for which the power consumption is estimated by using CPU performance counters and by measuring the number of disk accesses. In contrast to mobile computers, IoT devices are oftentimes built for a specific purpose. Therefore the energy required for certain services can be precisely measured during development, knowledge we use for our approach. The paper briefly mentions, without discussing any specific methods, that averaging and thresholding can be used for detecting processes responsible for fast battery depletion.  
In \cite{Buennemeyer2007a} the victim device measures its power consumption with a sampling rate of 10 kHz utilizing an additional power measurement circuit. For attack detection, the data is sent over the network to a trusted party which correlates the data with existing power traces indicating an attack. 
Compared to our approach this work has following disadvantages:
     1) it requires additional circuitry and the power traces must be sent over the network which may be challenging in constrained IoT networks.
     2) The detection algorithm needs reference power traces indicating an attack. How these traces are collected is not discussed. The authors clearly state their algorithm performs well when the attack causes long lasting high energy draw but performs poorly when the attack is conducted with spikes during usually inactive operational times. In contrast, our rate limitation algorithms perform well in both scenarios.
     3) \cite{Buennemeyer2007a} only considers attack detection but no countermeasures.
In \cite{Moyers2010, Buennemeyer2007b} a detection algorithm correlates the power traces also to Wi-Fi and Bluetooth activity for detecting other types of irregular behavior such as SYN flooding DoS, virus infections, or network probing. 

Further approaches for exhausting the batteries of IoT devices target weaknesses on the lower layers of the communication protocols. 
The authors of \cite{Brownfield2005, Raymond2006} analyze such attacks for several WSN MAC protocols and conclude that a protection mechanism has to consist of strong link-layer authentication, anti-replay protection, jamming identification and mitigation, and broadcast attack defense. 
Battery exhaustion attacks and countermeasures leveraging weaknesses in the routing layer were also considered in several publications \cite{Vasserman2013, Himabindhu2014}. The main idea of those attacks is that the attacker constructs packets which intentionally either circulate in routing loops exhausting each forwarding node or visit as many intermediate notes until they reach their destination.
This line of work remediates battery exhaustion leveraging lower-layer mechanisms, however leaves exhaustion through application-layer services out of consideration.

Note that, several methods to detect flooding DoS attacks on web servers were presented in previous work \cite{Sperotto2010, Wang2002, Siris2004, Dainotti2006}. A common characteristic is that these methods are designed to detect peak packet/request rates. Since a successful battery exhaustion attack does not necessarily exhibit peak rates (see \Cref{sec:detection}), but may for instance be conducted by increasing the request rate during usually less active times, we consider those methods poorly suitable for detecting battery exhaustion attacks.

\section{Conclusion}\label{sec:conclusion}
In this work, we considered IoT deployments where resource-constrained, battery-powered \Providers{} offer services 
that can be leveraged by malicious \Requesters{} for conducting battery exhaustion attacks.
We started with surveying representative use cases and deriving design goals for an effective countermeasure.
As a solution, we proposed a method combining rate limitation with lightweight authentication, supported by a trusted \Backend{}.
To the best of our knowledge we are the first who addressed the property of battery exhaustion attacks being feasible by requesting services at medium rates at usually less active operational times by using rate limitation.
For the rate limitation, we proposed the usage of two algorithms -- Leaky Bucket and EWMA -- and demonstrated their parametrization. 
Through simulation of a real word use case, we have shown that both are equally suitable for detecting such battery exhaustion attempts.
As part of our method, we proposed and formally verified two cryptographic authentication protocols suitable for different classes of IoT use cases.
We conducted our formal verification using the ProVerif verifier and made our verification models available online \cite{proverifModels}.
With our prototype implementation and evaluation, we have further shown that our method can be implemented in an energy-efficient way for battery-powered constrained devices, effectively reducing attack surface for battery exhaustion.

\bibliographystyle{IEEEtranS}
\interlinepenalty=10000
\bibliography{bibliography}

\end{document}